\documentclass[10pt,letterpaper]{article}

\usepackage{ccn}
\usepackage{pslatex}
\usepackage{apacite}
\usepackage{hyperref}
\usepackage{natbib}
\usepackage{lineno}
\usepackage{caption}
\usepackage{subcaption}

\usepackage{tikz}
\usetikzlibrary{positioning} 

\usepackage{amsmath, amssymb, natbib}
\usepackage{custom-definitions}

\title{Estimating Neural Representation Alignment \\ from Sparsely Sampled Inputs and Features}

\author{
{\large \bf Chanwoo Chun \thanks{These authors contributed equally to this work.}} \\
   Weill Cornell Medicine \\ Flatiron Institute
\And
{\large \bf Abdulkadir Canatar \footnotemark[1]} \\
   Flatiron Institute
 \And
{\large \bf SueYeon Chung} \\
   Flatiron Institute \\ New York University
 \And
{\large \bf Daniel D. Lee } \\
   Cornell Tech \\ Flatiron Institute
}

\begin{document}

\maketitle

\section{Abstract}
{
\bf
In both artificial and biological systems, the centered kernel alignment (CKA) has become a widely used tool for quantifying neural representation similarity. 
While current CKA estimators typically correct for the effects of finite stimuli sampling, the effects of sampling a subset of neurons are overlooked,
introducing notable bias in standard experimental scenarios. Here, we provide a theoretical analysis showing how this bias is affected by the representation geometry. We then introduce a novel estimator that corrects for both input and feature sampling.
We use our method for evaluating both brain-to-brain and model-to-brain alignments and show that it delivers reliable comparisons even with very sparsely sampled neurons. We perform within-animal and across-animal comparisons on electrophysiological data from visual cortical areas V1, V4, and IT data, and use these as benchmarks to evaluate model-to-brain alignment. 
We also apply our method to reveal how object representations become progressively disentangled across layers in both biological and artificial systems. These findings underscore the importance of correcting feature-sampling biases in CKA and demonstrate that our bias-corrected estimator provides a more faithful measure of representation alignment. The improved estimates increase our understanding of how neural activity is structured across both biological and artificial systems.
}
\begin{quote}
\small
\textbf{Keywords:} 
CKA; Representation Alignment; Disentanglement; Estimation
\end{quote}

\section{Introduction}
Over the past decade, the concept of \emph{representation similarity} has emerged as a powerful framework for comparing complex neural and computational systems \citep{kriegeskorte2008representational, kriegeskorte2013representational}. One of the most popular tools is \emph{centered kernel alignment} (CKA), a measure originally adapted from kernel-based independence metrics but now widely employed across fields such as machine learning, neuroscience, and cognitive science. In machine learning, CKA has become the de facto standard for quantifying how similarly different layers—or even entirely different architectures—encode the same input data \citep{kornblith2019similarity}. In neuroscience, CKA has emerged as a core analytical tool to assess whether neural populations, either within or across brain regions and species, produce similar activity patterns in response to identical stimuli \citep{yamins2016using, schrimpf2018brain}. Its popularity arises from two advantages: (1) CKA is invariant to orthogonal transformations, making it robust to small perturbations in feature space, and (2) it normalizes for overall variation in activity levels, facilitating meaningful comparisons based on alignment.

Despite these strengths, there is a growing consensus that conventional CKA estimators overlook a critical limitation in many experimental settings, that the sampling of only a subset of ``features” as with experimental recordings, can introduce a systematic bias. This is particularly significant in neuroscience, where only a fraction of the total neural population is recorded. Existing CKA estimators often assume that if enough data points (e.g., stimuli or input images) are provided, the measure becomes reliable. Yet this assumption ignores the additional requirement for sufficiently large samples along the ``feature” dimension. Consequently, researchers risk drawing misleading conclusions about the alignment of brain regions and neural network layers \citep{murphy2024correcting, cloos2024differentiable, han2023system, sucholutsky2023getting}.

In this work, we address this pressing concern. First, we show analytically how the geometry of high-dimensional representations contributes to spurious underestimation of similarity when only limited neuronal or model “units” are observed. Second, building on these insights, we introduce a novel estimator designed to remain consistent even with limited feature samples. By systematically correcting for finite-sample effects in both inputs and features, our method offers a more faithful gauge of representation alignment.

We demonstrate the real-world impact of our estimator through analyses of convolutional neural networks and multi-electrode electrophysiological recordings in visual cortices V1, V4, and IT \citep{papale2025extensive}. Our new estimator enables more accurate model-to-brain comparisons, revealing alignment trends that are otherwise corrupted by sampling biases. Beyond model-to-brain alignment, we show that the improved CKA estimator illuminates how object-category representations become increasingly disentangled along the primate ventral stream, similar to the observations in deep neural networks. Taken together, these results show that accounting for feature sampling is indispensable for robust representation analysis. Our work thus not only strengthens the theoretical foundations of CKA but also expands its practical utility for probing neural and computational representations.


\section{Problem Statement}
\begin{figure*}[t]
    \centering
    \includegraphics[width=1\textwidth]{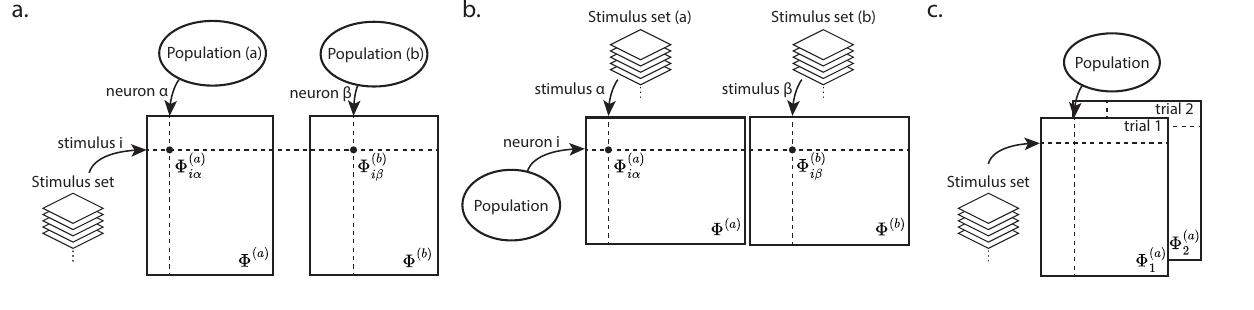}
    \caption{CKA can be considered in multiple different problem setups. Each rectangle represents a measurement matrix, and the dotted lines indicate specific rows and column indices. (a) Comparing the representations of two neural populations for one stimulus set. (b) Comparing the representations of two stimulus sets in one neural population. (c) Comparing the representations of a single population on a single stimulus set across two trials.}
    \label{fig:setups}
\end{figure*}

There are many scenarios in which the accurate quantification of the similarity between representations is of central interest (\cref{fig:setups}). For example, for a given stimulus set, one may want to compare the representations of the brain and a neural network model. To set a benchmark for brain-to-model similarity, it is also useful to measure brain-to-brain similarities across two individual animals or across trials in one animal. In a separate scenario, one may be interested in comparing the neural representations of two stimuli sets, e.g. two image categories, in a single brain region of one individual animal. In all these setups, CKA can be used to quantify the similarities.

Suppose one measures two separate neural representation matrices, $\X$ and $\Y$, where each row corresponds to a distinct stimulus and each column represents a recorded neuron. Here, assume the data has been preprocessed so that each neuron has zero mean across all the recorded stimuli.  Typically, in its most intuitive form, the CKA is then computed as

\begin{equation}\label{eq:basic}
\frac{\text{tr}(\X\X^\top \Y\Y^\top)}{\sqrt{\text{tr}\left(\left(\X\X^\top\right)^2\right)\text{tr}\left(\left(\Y\Y^\top\right)^2\right)  }}
\end{equation}
In a hypothetical scenario where one has access to all neurons and stimuli, CKA takes the value $1$ when the representations are perfectly aligned and $0$ when they span orthogonal subspaces. The normalization factor in the denominator keeps CKA invariant to the scaling of the representations.

In reality, the representation matrices are actually sampled submatrices of an unobserved larger matrix. Due to practical limitations, we often do not get to observe all neurons in a brain region, and it is not possible to present all possible stimuli from a stimulus set, e.g. natural images. Due to these sampling effects, computing the naive  CKA from the measurement matrices is heavily biased. Currently, the estimator derived from \cite{song2012feature} is widely used in the literature to correct the bias contributed by stimulus sampling. In this paper, we show how to correct the bias in the CKA estimator when both the stimuli and neurons are sampled from the underlying large matrix.

\section{Our Contribution}
We make the following contributions in this paper:
\begin{itemize}
    \item We find that the popular CKA estimator based on \cite{song2012feature} can take arbitrarily small values under finite sampling of both the stimuli and neurons, even when the underlying representations are perfectly aligned. Through theoretical analysis, we show that the bias increases with the intrinsic dimensionality of the representation. 
    
    \item To address this issue, we develop a more general CKA estimator that corrects the bias contributed by both the stimulus and neuron sampling. We demonstrate its reliability in both synthetic and neurophysiological data.
    
    \item We demonstrate our estimator enables a novel application of CKA that quantifies the representation disentanglement in the brain from real data. 
\end{itemize}

\section{Definitions}

\subsection{Centered Kernel Alignment}
A neural population $\bphi(x)$ defines a representation of stimuli $x \in \cX$ with a large number of neurons. The associated \textit{centered kernel function} $k(x,x')$ measures the similarity of two stimuli in the representation  space of $\bphi$ and is defined as:
\begin{align}\label{eq:centered_kernel_function}
    k(x, x') \coloneqq \Braket{\,\bphi(x) - \bE[\bphi]\;,\;\bphi(x') - \bE[\bphi]\,},
\end{align}
where $\braket{\cdot,\cdot}$ denotes an appropriate inner product. Here, $\bE[\bphi]$ denotes the expectation of $\bphi$ over the stimuli space $\cX$ and is used to center the population activity.

For two distinct neural populations $\bphi^{(a)}(x)$ and $\bphi^{(b)}(x)$, we measure their similarity using the Hilbert-Schmidt Independence Criterion (HSIC), a popular metric in machine learning \citep{gretton2005measuring}. HSIC compares two populations based on the correlations of their kernel functions $k^{(a)}$ and $k^{(b)}$, and yields, what we call an $\cH$-value:
\begin{align}\label{eq:hsic_definition}
    \cH(k^{(a)}, k^{(b)}) \coloneqq \bE_{x,x'}[k^{(a)}(x,x')k^{(b)}(x,x')].
\end{align}
A low $\cH$-value indicates less similarity and is zero if and only if the populations $\bphi^{(a)}$ and $\bphi^{(b)}$ vary independently.

The centered kernel alignment (CKA) is essentially the normalized version of HSIC \citep{cortes2012algorithms,kornblith2019similarity} and is defined as
\begin{align}\label{eq:cka_definition}
    \text{CKA}(k^{(a)}, k^{(b)}) \coloneqq \frac{\cH(k^{(a)}, k^{(b)})}{\sqrt{\cH(k^{(a)}, k^{(a)})\cH(k^{(b)}, k^{(b)})}}.
\end{align}
CKA is normalized to the interval $[0,1]$ and invariant to the overall magnitude of population activations.

\subsection{Measurement matrix}

Greek indices ($\alpha,\beta,\cdots$) denote neurons; Latin indices ($i,j,\cdots$) denote stimuli, and Latin indices ($a,b,\cdots$) denote distinct populations. We use the letter $Q$ for the number of neurons and $P$ for the number of stimuli. Quantities with a hat ($\;\widehat{\;}\;$)  indicate empirical estimates.

For each population $\bphi^{(a)}$, we observe $Q_a$ neurons $\bphi_{\alpha}(x)$ for $\alpha=1,\cdots,Q_a$. Each neuron is measured on the same set of stimuli $\{x_i\}$ for $i=1,\cdots,P$, where $P$ is the number of stimuli. The corresponding measurement matrix $\bPhi^{(a)} \in \bR^{P\times Q_a}$ with elements $\bPhi^{(a)}_{i\alpha} = \bphi^{(a)}_\alpha(x_i)$ denotes the response of each neuron to stimulus $x_i$.

We define the empirical \textit{uncentered} kernels (also called Gram matrices) by $$\K^{(a)} = \frac{1}{Q_a} \bPhi^{(a)}{\bPhi^{(a)}}^\top,$$ and the empirical centered kernels defined in \cref{eq:centered_kernel_function} by
\begin{align}
    \bar\K^{(a)} = \H\K^{(a)}\H, \quad \H = \I - \frac{1}{P}\1\1^\top,
\end{align}
where $\H$ is the centering matrix.

\section{Existing CKA estimators}


With these definitions, we now construct estimators of $\cH$-value from finite data. Here, we convey the overall ideas and provide detailed analyses in \cref{S:derive_est}.

\vspace{.5em}
\subsection{Naive Estimator}
Following the definition of $\cH$-value in \cref{eq:hsic_definition}, the naive estimator is
\begin{align}\label{eq:naive_estimator}
    \widehat{\cH}_0(k^{(a)}, k^{(b)}) & = \frac{1}{P^2} \sum_{i\neq j}\bar\K^{(a)}_{ij} \bar\K^{(b)}_{ji} + \frac{1}{P^2} \sum_{i}\bar\K^{(a)}_{ii} \bar\K^{(b)}_{ii},
\end{align}
which approaches the true $\cH$-value as $P\to\infty$. In this limit, the second term falls as $\cO(1/P)$ and hence does not contribute to the true $\cH$-value, demonstrating that the naive estimator $\widehat{\cH}_0$ for finite $P$ is biased\footnote{The centering operation is \textit{another} source of bias for $\widehat{\cH}_0$, which is discussed in \cref{S:bias}}. 

We define $\widehat{\text{CKA}}_0(k^{(a)}, k^{(b)})$ to be the naive estimator of CKA obtained by replacing $\widehat{\cH}_0$ in \cref{eq:cka_definition}. Observe that this naive estimator is biased due to the biases of $\widehat{\cH}_0$ which appears both in numerator and denominator.

\vspace{.5em}
\subsection{Stimulus-Corrected Estimator} The cause of the bias in the \cref{eq:naive_estimator} was the term where two independent stimuli indices $(i,j)$ coincided. In general, unbiased estimates are obtained by averaging over all indices where each stimulus is used at most once \citep{hoeffding1948class}. 

An unbiased estimator for $\cH$-value correcting for these terms was developed by \citet{song2012feature} and is given by
\begingroup\makeatletter\def\f@size{10}\check@mathfonts
\def\maketag@@@#1{\hbox{\m@th\large\normalfont#1}}%
\begin{align}
&\widehat{\cH}_{S}(k^{(a)},k^{(b)})
\coloneqq \frac{1}{P\left(P-3\right)}\\
&\times\left[\text{tr}\left(\tilde\K^{(a)}\tilde\K^{(b)}\right) -2\frac{\1^{\top}\tilde\K^{(a)}\tilde\K^{(b)}\1}{P-2} + \frac{\1^{\top}\tilde\K^{(a)}\1\1^{\top}\tilde\K^{(b)}\1}{\left(P-1\right)\left(P-2\right)}\right],\nonumber
\end{align}
\endgroup
where $\tilde \K_{ij} = \K_{ij}(1-\delta_{ij})$ is the uncentered kernel with diagonals removed. 

The sample corrected estimator for CKA, denoted by $\widehat{\text{CKA}}_S(k^{(a)}, k^{(b)})$, is obtained by replacing $\widehat{\cH}_{S}$ in \cref{eq:cka_definition}. This is the current version of CKA that is used both in deep learning \citep{nguyen2020wide,raghu2021vision,davari2022reliability} and neuroscience \citep{murphy2024correcting}.

Note that this estimator removes the bias due to coinciding stimuli indices. However, the inputs to $\widehat{\cH}_{S}(k^{(a)},k^{(b)})$ are kernels which involve a sum over neurons. Therefore, if its inputs have \textit{shared neurons} as in the case of the denominator of CKA, a similar bias discussed in \cref{eq:naive_estimator} occurs but for neuron indices. Therefore, assuming that the populations $\bphi^{(a)}$ and $\bphi^{(b)}$ have independent neurons, the numerator of $\widehat{\text{CKA}}_S(k^{(a)}, k^{(b)})$ remains unbiased, but its denominator is biased due to finite neuron sampling effects (See \cref{S:easybias}, \cref{eqn:numer_bias}).

\subsection{Bias of the Existing Estimator}

We find that the representation geometry affects the bias in the widely-used stimulus-corrected estimator $\widehat{\text{CKA}}_S$. Assuming the variance of the $\widehat{\cH}_S$ estimates are negligible, and the activation variance is normalized neuron-wise, the bias of $\widehat{\text{CKA}}_S$ can be approximated as
\begin{multline}\label{eq:s_bias}
    \bE \left[\widehat{\text{CKA}}_S(k^{(a)}, k^{(b)}) \right] -\text{CKA}(k^{(a)}, k^{(b)})\approx
    \\
    \left[
    \frac{1}{\sqrt{\left(1+\frac{\gamma_{a}-1}{Q_{a}}\right)\left(1+\frac{\gamma_{b}-1}{Q_{b}}\right)}}
    -1
    \right]\text{CKA}(k^{(a)}, k^{(b)})
\end{multline}
where $\gamma_a$ and $\gamma_b$ are the intrinsic dimensionalities of the two underlying representations, quantified by the participation ratio of the eigenvalues $\{\lambda_i\}$ of $\bar\K$ in the infinite data limit: $$\gamma=\frac{\left(\sum_i \lambda_i \right)^2}{\sum_i \lambda_i^2}.$$
From \cref{eq:s_bias}, we make the following observations:

\begin{itemize}
    \item $\widehat{\text{CKA}}_{S}$ always underestimates the true CKA on average.
    \item The scale by which $\widehat{\text{CKA}}_{S}$ underestimates is independent of the alignment but only dependent on the intrinsic dimensionalities and neuron sample sizes.
    \item Having larger intrinsic dimensionalities contributes to the greater underestimation of CKA and requires more neuron samples to mitigate the bias.
\end{itemize}

Therefore, if intrinsic dimensionality is much greater than the number of feature samples, $\gamma\gg  Q$, $\widehat{\text{CKA}}_S$ can take an arbitrarily small value even when two representations are perfectly aligned. On the other hand, our estimator is not affected by this issue. For more details, including the bias of $\widehat{\text{CKA}}_0$, see \cref{S:easybias}.

\vspace{.5em}
\section{Stimulus-Neuron-Corrected Estimator}Unlike the previous two estimators, an unbiased estimator correcting for both finite stimulus and neuron sampling must be a function of populations $\bphi^{(a)}$ rather than their kernels.

In this paper, we develop a novel estimator $\widehat{\cH}_C$ that corrects for finite neuron sampling effects when two populations have correlated, e.g., identical neurons. We leave its derivation to \cref{S:derive_est} and report the final estimator here
\begin{multline}
\widehat{\cH}_{C}(\bphi^{(a)},\bphi^{(b)}) = \frac{1}{P^{3}\left(P-3\right)Q_a\left(Q_b-\delta_{ab}\right)}\\
\times \sum_{ijlm}\left[c_{ijij}-\frac{2P}{P-2}\left(c_{ijjl}-\frac{c_{iiij}}{P}-\frac{c_{ijjj}}{P}+\frac{c_{iiii}}{2P}\right) \right.\\
\left.
+\frac{P^{2}}{\left(P-1\right)\left(P-2\right)}\left(c_{ijlm}-\frac{c_{iijl}}{P}-\frac{c_{jlii}}{P}+\frac{c_{iijj}}{P^{2}}\right)\right]
\end{multline}

where we introduced the tensor $c_{ijlm}$ defined as
\begingroup\makeatletter\def\f@size{9}\check@mathfonts\def\maketag@@@#1{\hbox{\m@th\large\normalfont#1}}%
\begin{align}
c_{ijlm}\coloneqq\sum_{\alpha\beta}\bPhi_{i\alpha}^{(a)}\bPhi_{j\alpha}^{(a)}&\bPhi_{l\beta}^{(b)}\bPhi_{m\beta}^{(b)}-\delta_{ab}\sum_{\alpha}\bPhi_{i\alpha}^{(a)}\bPhi_{j\alpha}^{(a)}\bPhi_{l\alpha}^{(b)}\bPhi_{m\alpha}^{(b)}.
\end{align}

\endgroup
The second term in $c_{ijlm}$ encodes the finite neuron correction only when two populations have identical neurons ($a=b$). Finally, we define our unbiased CKA estimator as
\begin{align}
\widehat{\text{CKA}}_C\left(\bphi^{(a)},\bphi^{(b)}\right) = \frac{\widehat{\cH}_{C}(\bphi^{(a)},\bphi^{(b)})}{\sqrt{\widehat{\cH}_{C}(\bphi^{(a)},\bphi^{(a)})\widehat{\cH}_{C}(\bphi^{(b)},\bphi^{(b)})}}.
\end{align}
Several remarks are in order:
\begin{itemize}
    \item Our estimator $\widehat{\cH}_C$ reduces to the stimulus-corrected estimator $\widehat{\cH}_S$ of \citet{song2012feature} when two populations $\bphi^{(a)}$ and $\bphi^{(b)}$ have independent neurons, generalizing previous results.

    \item When two distinct populations compared, our CKA estimator $\widehat{\text{CKA}}_C$ corrects the bias in the denominator of $\widehat{\text{CKA}}_S$ and their numerators remain identical. However, in certain cases, e.g. trial-to-trial similarity between two measurements of a single population, the numerator of $\widehat{\text{CKA}}_S$ also becomes biased.

    \item While the estimator looks complicated, it can be efficiently implemented in terms of \verb|einsum| operations. An implementation is provided in \cref{S:code}.
    \item Note that while the $\cH$-estimator we derive is unbiased, the proposed CKA-estimator is still biased due to non-linear operations (multiplication, square-root) involving $\widehat{\cH}_C$. Nevertheless, our CKA estimator is the least biased among the available estimators. Full analysis on this matter can be found in \cref{S:hardbias}. Also, as shown in the rest of the paper, we empirically find that the effect of this bias is much smaller compared to the effect of bias due to finite neuron sampling. One can also mitigate this bias by empirically averaging $\widehat{\mathcal{H}}$ over multiple trials and using that to estimate CKA. If $N$ estimates of $\widehat{\mathcal{H}}$ are available from independent trials, then this bias falls like $\mathcal{O}\left(1/N\right)$ (\cref{S:hardbias}). We will denote the CKA estimator derived from the empirical average of $N$ number of $\widehat{\mathcal{H}}$ by $\widehat{\text{CKA}^N}$.
\end{itemize}

\section{Linear example}

\begin{figure}[t]
    \centering
    \includegraphics[width=0.5\textwidth]{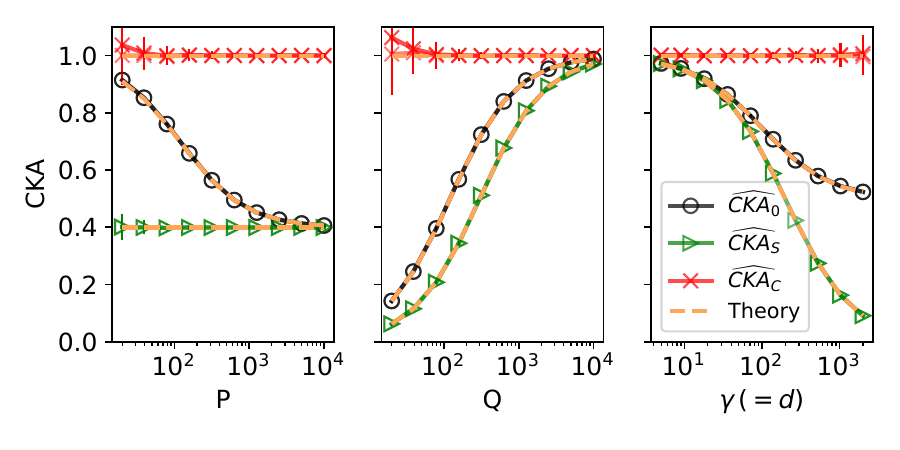}
    \caption{CKA estimators on the linear CKA example. The blue horizontal line is the true CKA. Left: Varying $P$, with $Q=200$ and $d=300$. Middle: Varying $Q$, with $P=200$ and $d=300$. Right: Varying the intrinsic dimensionality $\gamma$ which equals $d$ in this setup, with $P=Q=200$. The vertical error bar indicates the range of the first and third quartiles of the data (50\% of the data). The darker lines are $\widehat{\text{CKA}}$'s and the lighter lines are $\widehat{\text{CKA}^N}$'s where $N=500$. The yellow dotted lines are the theoretical predictions: for $\widehat{\text{CKA}}_C$, this is simply $1$; for $\widehat{\text{CKA}}_S$, we use \cref{eq:s_bias}; for $\widehat{\text{CKA}}_0$, we use \cref{Seq:o_bias}.}
    \label{fig:linear}
\end{figure}

Next, we numerically test all three estimators on a simple synthetic dataset with a known CKA value. We consider $d-$dimensional stimuli which are drawn from the distribution $\x \sim \cN(0, \I_{d})$. We define a linear population of the form $\bphi^{(a)}(\x) = \x^\top \w^{(a)}$, where each weight is drawn from the distribution $\w^{(a)} \sim \cN(0, \bSigma_a)$ and corresponds to a single neuron. Similarly, we define the population $\bphi^{(b)}(\x) = \x^\top \w^{(b)}$ where $\w^{(b)} \sim \cN(0, \bSigma_b)$. Since the entire stimuli and population distributions are known, the true CKA in \cref{eq:cka_definition} can be evaluated exactly: 
\begin{align}
    \text{CKA} = \frac{\tr (\bSigma_a \bSigma_b)}{\sqrt{\tr\bSigma_a^2 \tr\bSigma_b^2}}.
\end{align}
We consider $\bSigma_{a}=\bSigma_{b}= \I_{d}$ in which case both populations perfectly align with a $\text{CKA} = 1$. As indicated by \cref{eq:s_bias}, however, despite the perfect alignment of populations, the stimulus-corrected estimator $\widehat{\text{CKA}}_{S}$ can be arbitrarily small depending on the number of neurons sampled $Q$, and the dimensionality $\gamma$, which is exactly $d$ in this example. We numerically confirm these findings in \cref{fig:linear} and find that both estimators become highly sensitive when a limited number of neurons are observed. On the other hand, our estimator is able to recover the true CKA even at small neuron samples.

In this example, we observe that $\widehat{\text{CKA}^N}$'s and $\widehat{\text{CKA}}$'s almost completely overlap in all three estimators (\cref{fig:linear} lighter vs. darker lines), except when $Q$ is very small, $\widehat{\text{CKA}^N}_C$ and $\widehat{\text{CKA}}_C$ diverge a bit. This means that the bias correction for $\mathcal{H}$ contributes more to the bias correction of CKA than correcting the bias from the non-linear operations on the $\mathcal{H}$ estimates. The next section shows that this pattern is observed in neural data as well.

\begin{figure*}[ht!]
    \centering
    \includegraphics[width=1\textwidth]{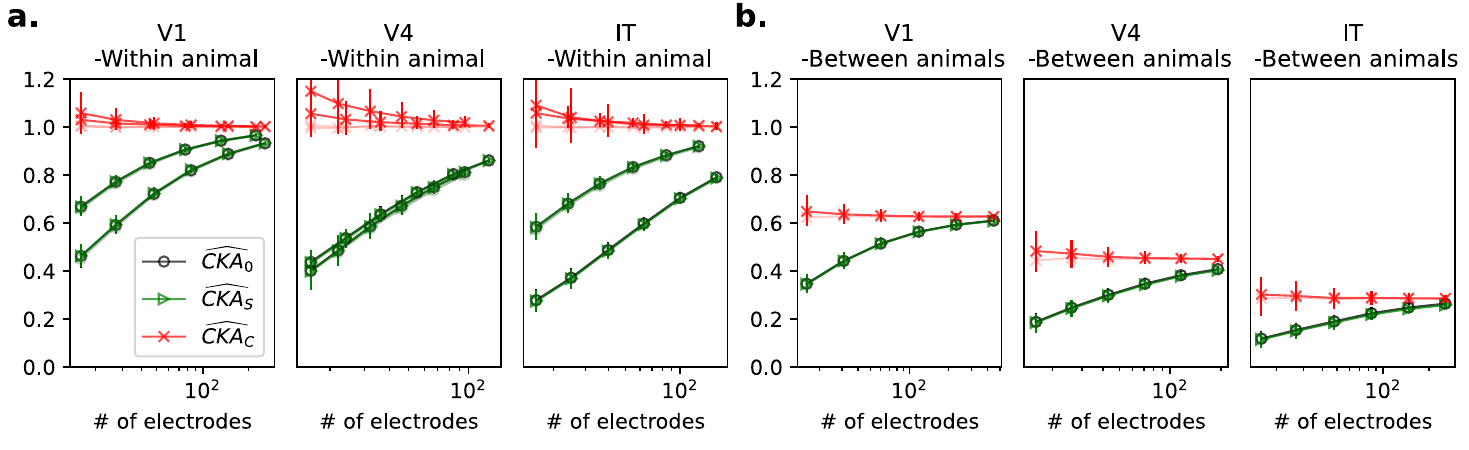}
    \caption{\textbf{a)} Each plot shows CKA values between disjoint sets of electrodes sampled from one brain region within one individual animal. The number of sampled electrodes ($Q$) is varied. One line for each animal. True CKA is 1. \textbf{b)}  Each plot shows CKA values between animals for one brain region. $\widehat{\text{CKA}}_{0}$ and $\widehat{\text{CKA}}_{S}$ are similar, since the number of stimuli is large: $P=2000$. The darker lines are $\widehat{\text{CKA}}$'s and the lighter lines are $\widehat{\text{CKA}^N}$'s where $N=1000$.}
    \label{fig:b2b}
\end{figure*}

\section{Practical applications in neuroscience}

In most of the brain recordings, the observed neurons are samples of much larger population. Here, we use electrophysiological data of the three key cortical regions for visual processing, V1, V4, and IT, in the order of processing cascade. It has been shown that the
neurons in V1 are simple filters, but the later regions V4 and IT are sensitive to semantic information \citep{hubel1962receptive,majaj2015simple,hung2005fast,cadena2024diverse}. \citet{papale2025extensive} presented 20,000 natural images of 1,854 object categories from the THINGS dataset \citep{hebart2019things} to two
monkeys and recorded neural responses with multiple electrode arrays over V1, V4, and IT, totaling 1046 electrodes in one monkey and 960 in the other. The measurement value is the average spiking voltage levels
of neurons adjacent to a given electrode averaged over a small time window immediately following an image presentation. We view the electrodes and images as rows and columns, respectively.

\subsection{Brain-to-Brain alignment}

Here use CKA to benchmark the similarity of a given brain region across individual animals (see \cref{fig:setups}a). Such brain-to-brain comparisons provide an essential reference for evaluating model-to-brain alignments. In our analysis, we first compare two disjoint sets of electrodes from the same brain region within a single animal. The ground truth in this within-region comparison is a perfect alignment (CKA = 1) on average. However, as shown in \cref{fig:b2b}a, both the naive estimator $\widehat{\text{CKA}}_{0}$ and the stimulus-corrected estimator $\widehat{\text{CKA}}_{S}$ significantly underestimate the true similarity, with their estimates strongly dependent on the number of neurons sampled. In contrast, our proposed estimator $\widehat{\text{CKA}}_{C}$ closely approximates the true value even with very limited neuron sampling, thereby highlighting the risk that conventional CKA estimates may fail to detect even perfect alignment when undersampling is present.

We further extend the analysis by measuring the similarity of each brain region across animals. As depicted in \cref{fig:b2b}b, $\widehat{\text{CKA}}_{C}$ yields reliable estimates over a range of neuron sampling sizes, whereas the conventional estimators remain highly sensitive to the sample size. Our results show that V1 representations are relatively conserved across individuals (with CKA around 0.6), IT representations exhibit lower similarity (around 0.3), and V4 lies in between (approximately 0.45). The similarity decreases over the layers of visual processing.

\subsection{Brain-to-Model alignment}

\begin{figure*}[ht!]
    \centering
    \includegraphics[width=\textwidth]{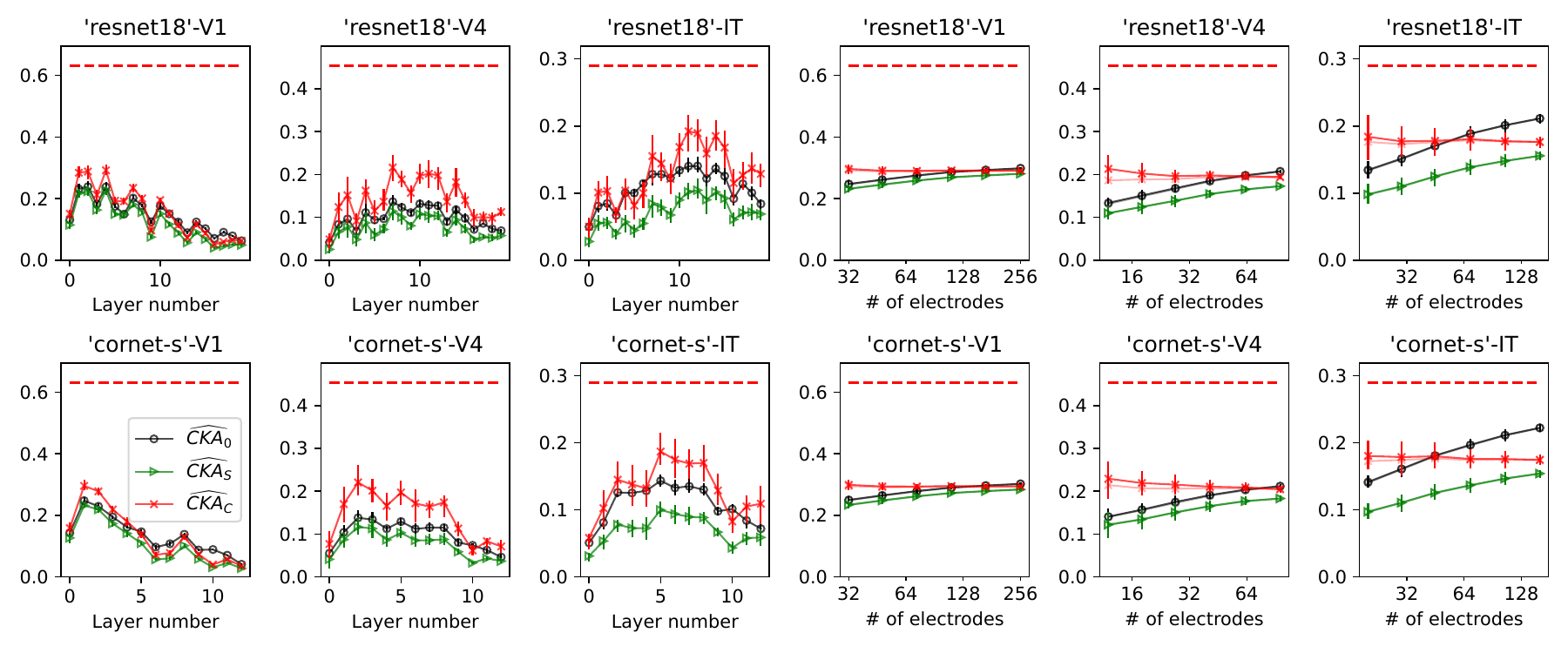}
    \caption{Comparison of the brain regions (V1,V4, and IT) and models (ResNet18 and CorNet-s) using CKA estimators. The left three columns show CKAs between each brain region and all model layers. The number of electrodes ($Q$) is downsampled to $1/16$ of all available electrodes in the dataset for each brain region.
    The right three columns show CKA between a brain region and the layer that is the most similar to the region. The number of electrodes ($Q$) varies from the factor of $1/16$ to $1$ (of all available electrodes) along the x-axis for each region. $P=1,000$ stimuli are used in all plots. The horizontal dotted line is the brain-to-brain $\widehat{\text{CKA}}_{C}$ across animals for each brain region from \cref{fig:b2b}b. The darker lines are $\widehat{\text{CKA}}$'s and the lighter lines are $\widehat{\text{CKA}^N}$'s, where $N=200$. The result for another monkey is shown in \cref{S:B2M}.}
    \label{fig:monkey_model}
\end{figure*}

The alignment between brain representations and artificial neural network layers is a topic of considerable interest. It has been repeatedly observed in the literature that the early visual regions, e.g., V1, have representations that are strikingly similar to the early layers of CNNs, and the later regions, e.g., IT, are aligned with the deeper layers \citep{yamins2014performance,schrimpf2018brain,nonaka2021brain}. As depicted in \cref{fig:setups}b, here we use the CKA estimators to compare the representations between brain regions (V1, V4, and IT) and layers of convolutional neural networks (CNNs), namely ResNet18 and CorNet-s \citep{he2016deep,kubilius2018cornet}. We then see how these CKA estimates compare against the brain-to-brain $\widehat{\text{CKA}}_{C}$ from the previous section.

We make the same observation of the V1-early layer alignment and IT-late layer alignment from all of the estimators. In the absolute scale, the V1-early layer alignment is generally higher, but the IT-late layer alignment is generally lower  (\cref{fig:monkey_model} left three plots). However, $\widehat{\text{CKA}}_{C}$ estimates that the alignment between animals also decreases by a similar factor. This indicates that in the relative scale, the model-to-brain alignments for all three regions are similarly close to the animal-to-animal alignments, an insight that cannot be reliably found with the other estimators.

We then test the sensitivity of the estimators to the number of electrodes. In \cref{fig:monkey_model} (right three plots), we pick the best alignment layer for each brain region and see how this alignment value changes with the number of electrodes. We observe that all estimators are similar for comparison with V1, meaning that there is only a small bias in $\widehat{\text{CKA}}_{0}$ and $\widehat{\text{CKA}}_{S}$. However, for the comparison with V4 and IT, the gap between the estimators is generally large, meaning we need our estimator $\widehat{\text{CKA}}_{C}$ for a reliable measurement. The $\widehat{\text{CKA}}_{0}$ and $\widehat{\text{CKA}}_{S}$ values are not converged even when all available electrodes are used.

\subsection{Object disentanglement}

Beyond inter-region comparisons, our estimator enables a novel application of CKA: quantifying the disentanglement of semantic information across object categories in the brain. Similar to the representation disentanglement observed in deeper layers of CNNs, the IT cortical region is believed to encode high-level semantic features that distinguish between objects \citep{yamins2014performance}. In this analysis, we estimate the CKA between pairs of object categories within a single brain region, where the columns of the measurement matrix correspond to stimulus images and the rows correspond to the shared neurons (as illustrated in \cref{fig:setups}b). A lower CKA in this context suggests greater semantic disentanglement.

\begin{figure*}[t]
    \centering
    \includegraphics[width=\textwidth]{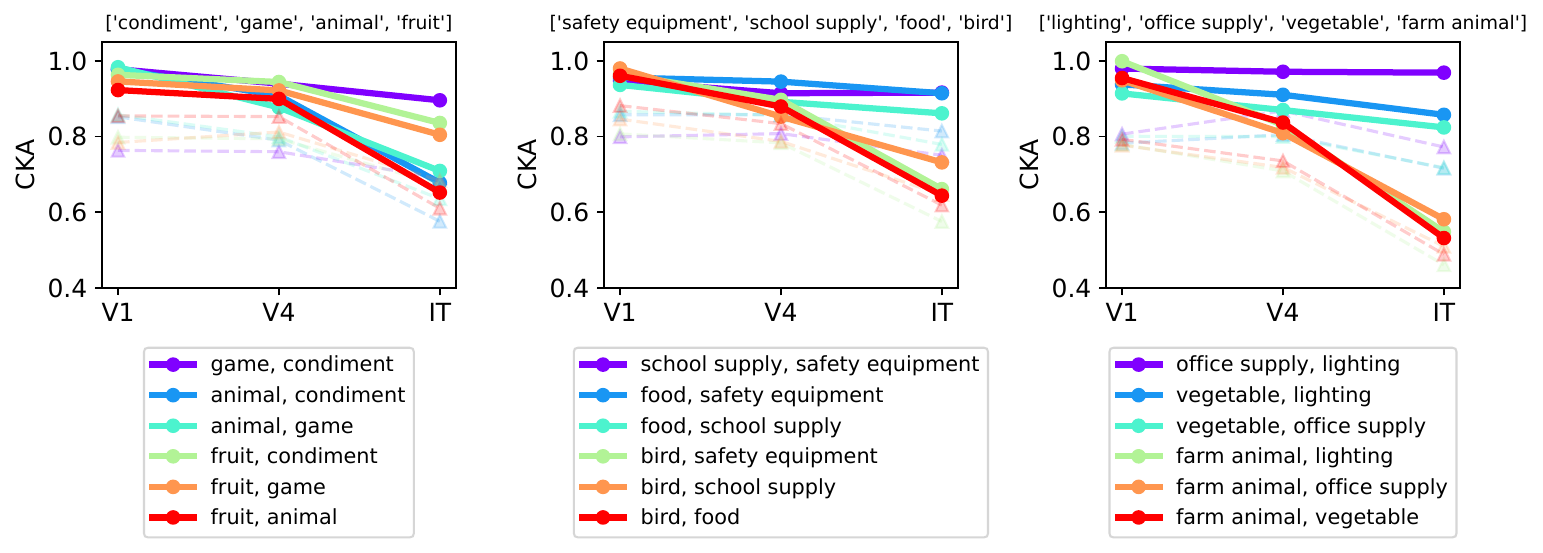}
    \caption{Quantifying semantic disentanglement in the brain with CKA. Solid line represents $\widehat{\text{CKA}}_{C}$, whereas dotted line represents $\widehat{\text{CKA}}_{S}$. Three separate groups of object images were prepared, and all pairwise comparisons were performed in each group. The result for another monkey is shown in the \cref{S:OD}.}
    \label{fig:obj}
\end{figure*}

We observe that the estimated CKA between object categories generally decreases over V1, V4, and IT, indicating a gradual semantic disentanglement over these regions (\cref{fig:obj}). Interestingly, natural object pairs get separated faster than artificial object pairs. This pattern is more strongly observed in $\widehat{\text{CKA}}_{C}$ than in $\widehat{\text{CKA}}_{S}$. Interestingly, there are many cases where $\widehat{\text{CKA}}_{S}$ indicates high disentanglement (small CKA) but ours $\widehat{\text{CKA}}_{C}$ indicates low disentanglement (large CKA). This is observed in V1 for all pairwise comparisons: Our estimator consistently estimates CKA value near 1, suggesting V1 does not encode semantic information, whereas the heavily biased $\widehat{\text{CKA}}_{S}$ spuriously suggests otherwise. Also, in IT, $\widehat{\text{CKA}}_{C}$ is near 1 for artificial objects pairs, but the bias in $\widehat{\text{CKA}}_{S}$ spuriously suggests the objects are disentangled.

\section{Trial-to-Trial Similarity}

\begin{figure*}[ht!]
    \centering
    \begin{subfigure}{.5\textwidth}
      \centering
      \includegraphics[width=.9\linewidth]{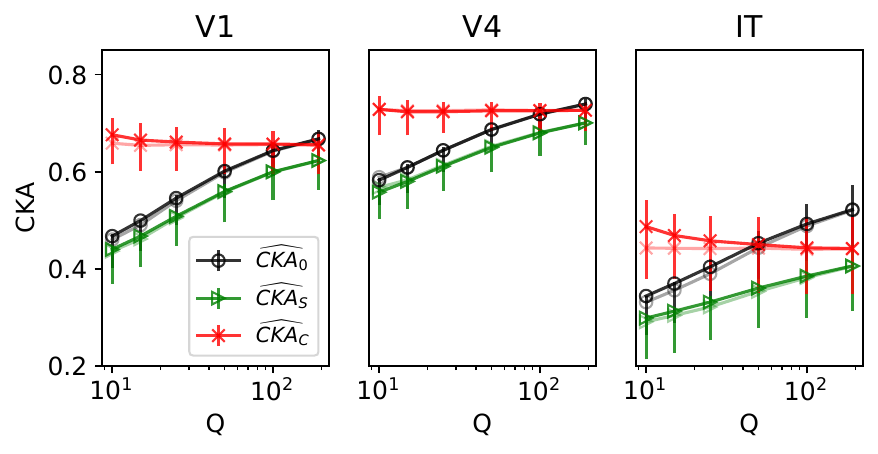}
      \caption*{Animal 1}
      \label{fig:tt0}
    \end{subfigure}%
    \begin{subfigure}{.5\textwidth}
      \centering
      \includegraphics[width=.9\linewidth]{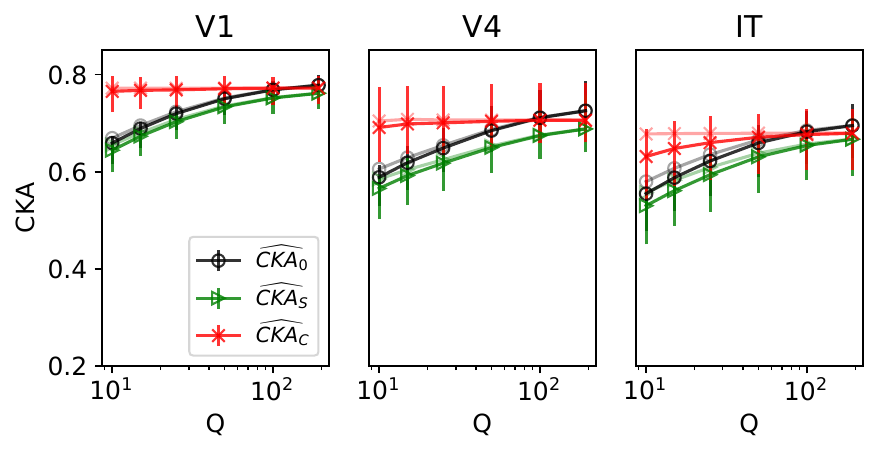}
      \caption*{Animal 2}
      \label{fig:tt1}
    \end{subfigure}%
    \caption{Trial-to-trial CKA estimates of brain regions V1, V4 and IT as a function of neuron sample size $Q$. Measurements on two animals are shown separately in the left plot and right plot. The darker lines are $\widehat{\text{CKA}}$'s and the lighter lines are $\widehat{\text{CKA}^N}$'s.}
    \label{fig:trial}
\end{figure*}

As our final application, we evaluate the trial-to-trial similarity of neural recordings from the same brain region on the same stimuli set as depicted in \cref{fig:setups}c. We use the neural recordings from \cite{papale2025extensive} of visual cortical areas V1, V2, and IT over $30$ trials on $100$ images. This case differs from the previous analyses since a single population $\bphi^{(a)}$ is compared across two trials ($\bphi^{(a)}_1$ and $\bphi^{(a)}_2$). In this case, $\widehat{\text{CKA}}_{S}$ becomes biased in its numerator as well as its denominator, which our estimator accounts for correctly.

In \cref{fig:trial}, we calculate the CKA between all pair-wise single trial measurements for each brain region and report the mean of each estimator. As before, our estimator yields consistent estimates across different neuron samplings $Q$, while the others remain highly biased. Furthermore, we observe that trial-to-trial similarities for regions V1 and V4 are significantly higher than the region IT. This indicates that the trial-to-trial variability in IT is larger than in earlier visual regions.

\section{Discussion}
 \cite{murphy2024correcting} also highlights the sensitivity of the naive CKA estimator to the number of features, with a focus on the ratio of the number of inputs and features. They motivate the problem by showing that the naive CKA (\cref{eq:basic}) between two independent random matrices $X$ and $Y$ (their entries are i.i.d. standard normal) takes a non-zero value which depends on the ratio of the number of stimuli and neurons:
 \begin{equation}\label{eq:random}
     \frac{1}{\sqrt{\left(1+\frac{P}{Q_a}\right)\left(1+\frac{P}{Q_b}\right)}}
 \end{equation}
in the limit of large $P$, $Q_a$, and $Q_b$. They observe that $\widehat{\text{CKA}}_{S}$ takes value $0$ in the same setup, resolving this issue of having spurious similarities between independent random matrices. While this observation is correct, \cite{murphy2024correcting} interprets this result as $\widehat{\text{CKA}}_{S}$ resolving the issue of the naive CKA being sensitive to $Q_a$ and $Q_b$ (denoted by $P_1$ and $P_2$ in their paper), since $\widehat{\text{CKA}}_{S}$ value is $0$ regardless of $Q_a$ and $Q_b$. Extrapolating this interpretation, they apply $\widehat{\text{CKA}}_{S}$ to real neural data in an attempt to resolve the issue of CKA sensitivity to the number of neurons.

However, here we explain that this interpretation is only valid when the true CKA is 0, and therefore cannot be generalized to other cases. $\widehat{\text{CKA}}_{S}$ returns $0$ in the random matrix setup, not because it corrects neuron sampling, but because it corrects the stimulus sampling. We can see that by taking the number of stimuli to infinity ($P\to \infty$) in \cref{eq:random}, which makes the CKA estimate approach $0$. In practical scenarios, where the measurement matrices are not random, $\widehat{\text{CKA}}_{S}$ is still sensitive to the number of features as we have shown in \cref{eq:s_bias}. $\widehat{\text{CKA}}_{C}$ presented in this paper resolves this problem.

Finally, our estimator for the $\cH$-value can be used for de-biasing other HSIC-based CKA measures, such as angular-CKA introduced in \cite{williams2021generalized} and Representational Similarity Analysis (RSA) \citep{kriegeskorte2008representational}, since these measures also ignore the bias coming from finite neuron sampling.

\section{Conclusion}

We have addressed a key limitation in applying CKA to scenarios where only a subset of features—such as neurons or model units—is observable. While CKA is widely used to compare representations in machine learning and neuroscience, existing estimators systematically underestimate similarity when features are undersampled, a common issue in neural data collection or reduced-dimension network analyses.

Theoretically, we showed how the stimulus-corrected $\mathcal{H}$ estimator still fails to yield unbiased CKA under partial column sampling. We derived a bias-corrected estimator that handles both input and feature sampling, more accurately recovering true representation similarity. Empirically, our method performed well on both synthetic data and electrophysiological recordings from the ventral visual stream (V1, V4, and IT). Conventional approaches underestimated alignment, particularly for high-dimensional or semantically complex representations, whereas our estimator revealed consistent neural similarity patterns and clearer evidence of object-category disentanglement.


\bibliographystyle{ccn_style}
\bibliography{ccn_style}

\newpage
\renewcommand{\theequation}{S\arabic{equation}}
\renewcommand{\thefigure}{S\arabic{figure}}
\setcounter{equation}{0}
\setcounter{figure}{0}

\appendix
\setcounter{secnumdepth}{2}
\onecolumn

\section*{\Huge{Supplementary Information}}

\section{Full derivation of the \texorpdfstring{$\cH$}{H} estimators}\label{S:derive_est}

\subsection{Expression of \texorpdfstring{$\cH$}{H} in terms of the uncentered kernel}

Recall that the kernel is defined as $k^ {}(x,y)=\left\langle \phi(x),\phi(y)\right\rangle $
and the centered kernel is $k'(x,y)=\left\langle \phi(x)-\mathbb{E}_{x}\left[\phi(x)\right],\phi(y)-\mathbb{E}_{x}\left[\phi(x)\right]\right\rangle $.
Then $\mathcal{H}$ is defined as

\[
\mathcal{H}(k^{(a)},k^{(b)})=\mathbb{E}_{x,y}\left[k'{}^{(a)}(x,y)k'{}^{(b)}(x,y)\right]
\]

Note that $k'$ can be written in terms of $k$:

\[
k'(x,y)=\left\langle \phi(x)-\mathbb{E}_{x}\left[\phi(x)\right],\phi(y)-\mathbb{E}_{x}\left[\phi(x)\right]\right\rangle 
\]

\[
=k(x,y)-\mathbb{E}_{z}\left[k(x,z)\right]-\mathbb{E}_{z}\left[k(z,y)\right]+\mathbb{E}_{z,w}\left[k(z,w)\right].
\]
Plugging this into our definition of $\mathcal{H}$, we arrive at the expression
of $\mathcal{H}$ explicitly in terms of the original uncentered kernels, $k^{(a)}$
and $k^{(b)}$:

\[
\mathcal{H}(k^{(a)},k^{(b)})=\mathbb{E}_{x,y}\left[k^{(a)}(x,y)k^{(b)}(x,y)\right]-2\mathbb{E}_{x,y,z}\left[k^{(a)}(x,y)k^{(b)}(x,z)\right]+\mathbb{E}_{x,y}\left[k^{(a)}(x,y)\right]\mathbb{E}_{x,y}\left[k^{(b)}(x,y)\right]
\]

We want to derive an estimator where each term has an expected value
that is equal to an individual term above.

\subsection{Identifying the source of bias in the naive estimator}

We begin our analysis by considering the naive $\mathcal{H}$ estimator:

\[
\widehat{\mathcal{H}}_{0}(\Phi^{(a)},\Phi^{(b)})=\frac{1}{P^{2}}\text{tr}\left(HK^{(a)}HK^{(b)}\right)
\]
where $H=I-\frac{1}{P}11^{\top}$ and $K^{(a)}=\frac{1}{Q}\Phi^{(a)}\Phi^{(a)\top}$.
When we expand this, we get

\[
\widehat{\mathcal{H}}_{0}(\Phi^{(a)},\Phi^{(b)})=\frac{1}{P^{2}}\sum_{ij=1}^{P}K_{ij}^{(a)}K_{ij}^{(b)}-\frac{2}{P^{3}}\sum_{ijl=1}^{P}K_{ij}^{(a)}K_{jl}^{(b)}+\frac{1}{P^{4}}\sum_{ijlm=1}^{P}K_{ij}^{(a)}K_{lm}^{(b)}.
\]

Note that $K^{(a)}$ and $K^{(b)}$ are dependent on the stimuli $\left\{ x_{i}\right\} _{i=1}^{P}$
such that, for example $K_{ij}^{(a)}=k^{(a)}(x_{i},x_{j})$. Now,
let us take the average of $\widehat{\mathcal{H}}_{0}$ over all possible
stimuli sets $\left\{ x_{i}\right\} _{i=1}^{P}$ sampled from $\mathcal{X}$.
We can compute the expected value separately for each term and add
them together for the final result. Let us consider the first term:

\[
\mathbb{E}_{\left\{ x_{i}\right\} _{i=1}^{P}}\left[\frac{1}{P^{2}}\sum_{ij=1}^{P}K_{ij}^{(a)}K_{ij}^{(b)}\right]=\frac{1}{P^{2}}\mathbb{E}_{\left\{ x_{i}\right\} _{i=1}^{P}}\left[\sum_{ij=1}^{P}k^{(a)}(x_{i},x_{j})k^{(b)}(x_{i},x_{j})\right]
\]

\[
=\frac{1}{P^{2}}\sum_{i\neq j}^{P}\mathbb{E}_{x,x'}\left[k^{(a)}(x,x')k^{(b)}(x,x')\right]+\frac{1}{P^{2}}\sum_{i}^{P}\mathbb{E}_{x}\left[k^{(a)}(x,x)k^{(b)}(x,x)\right]
\]

\[
=\left(\frac{P-1}{P}\right)\mathbb{E}_{x,x'}\left[k^{(a)}(x,x')k^{(b)}(x,x')\right]+\frac{1}{P}\mathbb{E}_{x}\left[k^{(a)}(x,x)k^{(b)}(x,x)\right]
\]

\[
=\mathbb{E}_{x,x'}\left[k^{(a)}(x,x')k^{(b)}(x,x')\right]+\frac{1}{P}\left(\mathbb{E}_{x}\left[k^{(a)}(x,x)k^{(b)}(x,x)\right]-\mathbb{E}_{x,x'}\left[k^{(a)}(x,x')k^{(b)}(x,x')\right]\right)
\]
Notice that in the second equality, we separate the sum into two parts,
based on whether the indices overlap or not: $i=j$ vs $i\neq j$.
This is essential since the expected values are different for these
two cases. Note that in the second term here is $\mathcal{O}(1/P)$,
contributing as bias. We would not have gotten this bias if the term
$\frac{1}{P^{2}}\sum_{ij=1}^{P}K_{ij}^{(a)}K_{ij}^{(b)}$ was instead
defined as $\frac{1}{P\left(P-1\right)}\sum_{i\neq j}^{P}K_{ij}^{(a)}K_{ij}^{(b)}$.
In that case, the expected value does not have a bias:

\[
\mathbb{E}_{\left\{ x_{i}\right\} _{i=1}^{P}}\left[\frac{1}{P\left(P-1\right)}\sum_{i\neq j}^{P}K_{ij}^{(a)}K_{ij}^{(b)}\right]=\mathbb{E}_{x,x'}\left[k^{(a)}(x,x')k^{(b)}(x,x')\right].
\]
Note that $\frac{1}{P\left(P-1\right)}$ is introduced as a scaling
factor since $P\left(P-1\right)$ is the number of summand in $\sum_{i\neq j}^{P}$.

\subsection{Derivation of the stimulus-corrected \texorpdfstring{$\cH$}{H} estimator}

Applying the same logic to the rest of the terms in $\widehat{\mathcal{H}}_{0}(\Phi^{(a)},\Phi^{(b)})$,
we arrive at an estimator that removes the bias from input sampling:

\[
\widehat{\mathcal{H}}_{S}(\Phi^{(a)},\Phi^{(b)})=\frac{1}{P\left(P-1\right)}\sum_{i\neq j}^{P}K_{ij}^{(a)}K_{ij}^{(b)}-\frac{2}{P\left(P-1\right)\left(P-2\right)}\sum_{i\neq j\neq l}^{P}K_{ij}^{(a)}K_{jl}^{(b)}+\frac{1}{P\left(P-1\right)\left(P-2\right)\left(P-3\right)}\sum_{i\neq j\neq l\neq m}^{P}K_{ij}^{(a)}K_{lm}^{(b)}.
\]
This is equivalent to the estimator by Song et al. Note that these
sums over disjoint indices, e.g. $\sum_{i\neq j\neq l}^{P}$, are
practically difficult to compute, so they are decomposed into a linear
combination of regular sums. Denoting $K_{ij}^{(a)}K_{lm}^{(b)}$
as $v_{ijlm}$, the first term can be rewritten as:

\[
\sum_{i\neq j}^{P}K_{ij}^{(a)}K_{ij}^{(b)}=\sum_{ij}v_{ijij}-\sum_{i}v_{iiii},
\]
whereas the second term can be:

\[
\sum_{i\neq j\neq l}^{P}K_{ij}^{(a)}K_{jl}^{(b)}=\sum_{ijl}v_{ijjl}-\sum_{ij}v_{iiij}-\sum_{ij}v_{ijjj}-\sum_{ij}v_{ijij}+\sum_{i}2v_{iiii},
\]
and the third term:

\[
\sum_{i\neq j\neq l\neq m}^{P}K_{ij}^{(a)}K_{lm}^{(b)}=\sum_{ijlm}v_{ijlm}-\sum_{ijl}\left(v_{iijl}+v_{jlii}+4v_{ijjl}\right)+\sum_{ij}\left(v_{iijj}+4v_{iiij}+4v_{ijjj}+2v_{ijij}\right)-\sum_{i}6v_{iiii}
\]

Replacing the terms in $\widehat{\mathcal{H}}_{S}(\Phi^{(a)},\Phi^{(b)})$
with these new notations, we arrive at the following expression:

\begin{multline}
\widehat{\mathcal{H}}_{S}(\Phi^{(a)},\Phi^{(b)})=
\frac{1}{P\left(P-3\right)}\times\\
\left(\left(\sum_{ij}v_{ijij}-\sum_{i}v_{iiii}\right)-\frac{2}{\left(P-2\right)}\left(\sum_{ijl}v_{ijjl}-\sum_{ij}v_{iiij}-\sum_{ij}v_{ijjj}+\sum_{i}v_{iiii}\right)+\frac{1}{\left(P-1\right)\left(P-2\right)}\left(\sum_{ijlm}v_{ijlm}-\sum_{ijl}v_{iijl}-\sum_{ijl}v_{jlii}+\sum_{ij}v_{iijj}\right)\right)    
\end{multline}

\begin{multline}
\widehat{\mathcal{H}}_{S}(\Phi^{(a)},\Phi^{(b)})=\frac{1}{P^{3}\left(P-3\right)}\times 
\\
\sum_{ijlm}\left(\left(v_{ijij}-\frac{v_{iiii}}{P}\right)-\frac{2P}{P-2}\left(v_{ijjl}-\frac{v_{iiij}}{P}-\frac{v_{ijjj}}{P}+\frac{v_{iiii}}{P^{2}}\right)+\frac{P^{2}}{\left(P-1\right)\left(P-2\right)}\left(v_{ijlm}-\frac{v_{iijl}}{P}-\frac{v_{jlii}}{P}+\frac{v_{iijj}}{P^{2}}\right)\right) 
\end{multline}

Suppose $K'$ is a version of $K$ whose diagonal elements are $0$,
and $v'_{ijlm}\coloneqq K'{}_{ij}^{(a)}K'{}_{lm}^{(b)}$. Then, the
above expression simplifies to 

\[
\widehat{\mathcal{H}}_{S}(\Phi^{(a)},\Phi^{(b)})=\frac{1}{P\left(P-3\right)}\left(\sum_{ij}v'_{ijij}-\frac{2}{\left(P-2\right)}\sum_{ijl}v'_{ijjl}+\frac{1}{\left(P-1\right)\left(P-2\right)}\sum_{ijlm}v'_{ijlm}\right)
\]

\[
=\frac{1}{P\left(P-3\right)}\left(\text{tr}\left(K'{}^{(a)}K'{}^{(b)}\right)-\frac{2}{\left(P-2\right)}1^{\top}K'{}^{(a)}K'{}^{(b)}1+\frac{1}{\left(P-1\right)\left(P-2\right)}1^{\top}K'{}^{(a)}11^{\top}K'{}^{(b)}1\right)
\]
which is the exact expression found in \cite{song2012feature}.

\subsection{Derivation of the stimulus-neuron-corrected \texorpdfstring{$\cH$}{H} estimator}

Our estimator assumes that the features are also sampled, and the features are correlated or identical in (a) and (b). Let us consider
a single term $v_{ijlm}$ (redefined here as $K_{ij}^{(a)}K_{lm}^{(a)}$ reflecting $a=b$) and see how the sampling of features contribute
to the bias.

\[
v_{ijlm}=K_{ij}^{(a)}K_{lm}^{(a)}=\frac{1}{Q^{2}}\sum_{\alpha\beta}\Phi_{i\alpha}^{(a)}\Phi_{j\alpha}^{(a)}\Phi_{l\beta}^{(a)}\Phi_{m\beta}^{(a)}
\]
If we were to take the expected value of $v_{ijlm}$ over the feature
sampling, we need to consider the overlapping of the feature indices
$\alpha$ and $\beta$. Therefore, the following should be an unbiased
estimator of $v_{ijlm}$:

\[
c'_{ijlm}=\frac{1}{Q\left(Q-1\right)}\sum_{\alpha\neq\beta}\Phi_{i\alpha}^{(a)}\Phi_{j\alpha}^{(a)}\Phi_{l\beta}^{(a)}\Phi_{m\beta}^{(a)}=\frac{1}{Q\left(Q-1\right)}\left(\sum_{\alpha\beta}\Phi_{i\alpha}^{(a)}\Phi_{j\alpha}^{(a)}\Phi_{l\beta}^{(a)}\Phi_{m\beta}^{(a)}-\sum_{\alpha}\Phi_{i\alpha}^{(a)}\Phi_{j\alpha}^{(a)}\Phi_{l\alpha}^{(a)}\Phi_{m\alpha}^{(a)}\right).
\]
In the main text we define $c_{ijlm}$, which is simply $c'_{ijlm}$
without the $\frac{1}{Q\left(Q-1\right)}$ factor, i.e. $c_{ijlm}=Q\left(Q-1\right)c'_{ijlm}$.
Therefore, finally, our estimator can be expressed as

\begin{multline}
\widehat{\mathcal{H}}_{C}(\Phi^{(a)},\Phi^{(a)})=\frac{1}{P^{3}\left(P-3\right)Q\left(Q-1\right)}\times
\\
\sum_{ijlm}\left(\left(c_{ijij}-\frac{c_{iiii}}{P}\right)-\frac{2P}{P-2}\left(c_{ijjl}-\frac{c_{iiij}}{P}-\frac{c_{ijjj}}{P}+\frac{c_{iiii}}{P^{2}}\right)+\frac{P^{2}}{\left(P-1\right)\left(P-2\right)}\left(c_{ijlm}-\frac{c_{iijl}}{P}-\frac{c_{jlii}}{P}+\frac{c_{iijj}}{P^{2}}\right)\right)    
\end{multline}

\section{Bias analysis}\label{S:bias}
\subsection{Generative process framework}
Here we formalize the problem setup by defining a general formulation
of the process that generates the measurement matrices. Consider a
pair of measurement matrices $\Phi_{(a)}$ and $\Phi_{(b)}$ of two
systems $a$ and $b$. Let $x_{i}\in\mathcal{X}$ 
be latent variables for the $i$th rows of $\Phi_{(a)}$ and $\Phi_{(b)}$, and $u_{\alpha}\in\mathcal{U}$ and $v_{\alpha}\in\mathcal{V}$
be column-latent variables for the $\alpha$th column of $\Phi_{(a)}$,
and $\Phi_{(b)}$, respectively. Let $P$ be the number of row-latent variables sampled, and $Q_a$ and $Q_b$ be the numbers of column-latent variables sampled. Let $\phi_{a}:\mathcal{X}\times\mathcal{U}\to\mathbb{R}$
be a map that defines the measurement value. Then we assume the entries
of $\Phi_{(a)}\in\mathbb{R}^{P\times Q_{a}}$ and $\Phi_{(b)}\in\mathbb{R}^{P\times Q_{b}}$
are defined as
\[
\Phi_{i\alpha}^{(a)}=\phi_{a}(x_{i},u_{\alpha}),\text{and }\Phi_{i\beta}^{(b)}=\phi_{b}(x_{i},v_{\beta}).
\]
In some cases, a latent space might be fully observed in the measurement.
For example, $N$ dimensional layer of neural network activation
is given by a feature map $\psi_{(a)}:\mathcal{X}\to\mathbb{R}^{N}$,
where $\mathcal{X}$ is the input space, and the column latent set
$\mathcal{U}$ of cardinality $N$ is fully observed. Here, $\mathcal{U}$
would be a set of trained neural network weights. The uncentered kernel would be simply
$k'(x,x')=\frac{1}{N}\psi(x)\psi(x')^{\top}$.
If the latent variables cannot be fully observed, we assume there
are probability measures over the latent spaces: $\rho_{\mathcal{X}}$
is the probability measures over $\mathcal{X}$
and, and similarly for $\rho_{\mathcal{U}}$ and $\rho_{\mathcal{V}}$.
Assume $\phi_{a}$ is square integrable w.r.t. $\rho_{\mathcal{X}}$
and $\rho_{\mathcal{U}}$, and similarly $\phi_{b}$ is also square
integrable. The associated uncentered kernels are defined as $k'_{a}(x,x')=\int d\rho_{\mathcal{U}}(u)\:\phi_{a}(x,u)\phi_{a}(x',u)$
and $k'_{b}(y,y')=\int d\rho_{\mathcal{V}}(v)\:\phi_{b}(y,v)\phi_{b}(y',v)$. We may also define the associated kernel integral operator:

\begin{equation}
    T_kf=\int d\rho_\mathcal{X}(x)\, k'(\cdot,x)f(x).
\end{equation}
Later, the eigenvalues of the operator $T_k$ will be relevant.

We also define associated covariance kernels:
$\tilde{k}'_{a}(u,u')=\int d\rho_{\mathcal{X}}(x)\:\phi_{a}(x,u)\phi_{a}(x,u')$ and $\tilde{k}'_{b}(v,v')=\int d\rho_{\mathcal{X}}(x)\:\phi_{b}(x,v)\phi_{b}(x,v')$.

\subsection{Biases in CKA estimators contributed by the biases in HSIC estimators}\label{S:easybias}

Here we derive the analytical expression of the biases of the CKA
estimators. Here we assume that each feature is already centered,
i.e. $\int d\rho_{\mathcal{X}}(x)\:\phi_{a}(x,\cdot)=0$, and $\int d\rho_{\mathcal{X}}(x)\:\phi_{b}(x,\cdot)=0$
. We first compute the expected values of the HSIC estimators. Let
$\mathcal{S}_{\mathcal{X}}\coloneqq\left\{ x_{i}\right\} _{i=1}^{P}$,
$\mathcal{F}_{\mathcal{U}}\coloneqq\left\{ u_{i}\right\} _{i=1}^{Q_{a}}$,
and $\mathcal{F}_{\mathcal{V}}\coloneqq\left\{ v_{i}\right\} _{i=1}^{Q_{b}}$
be the sets of latent variables sampled independently from $\rho_{\mathcal{X}},$
$\rho_{\mathcal{U}}$, and $\rho_{\mathcal{V}}$ respectively. The
naive HSIC estimator is then given by

\[
\widehat{\mathcal{H}_{0}}(\phi_{a},\phi_{b},\mathcal{S}_{\mathcal{X}},\mathcal{F}_{\mathcal{U}},\mathcal{F}_{\mathcal{V}})=\frac{1}{P^{2}Q_{a}Q_{b}}\sum_{ij}\sum_{\alpha}\sum_{\beta}\phi_{a}(x_{i},u_{\alpha})\phi_{a}(x_{j},u_{\alpha})\phi_{b}(x_{i},v_{\beta})\phi_{b}(x_{j},v_{\beta})
\]
if $u_{\alpha}$ and $v_{\beta}$ are independently sampled for all
$\alpha$ and $\beta$ combinations, which corresponds to the numerator
of CKA. If the column latent variables are identical, and $\phi_{a}=\phi_{b}$,
this corresponds to the HSICs in the denominator of naive CKA:
\[
\widehat{\mathcal{H}_{0}}(\phi_{a},\mathcal{S}_{\mathcal{X}},\mathcal{F}_{\mathcal{U}})=\frac{1}{P^{2}Q_{a}^{2}}\sum_{ij}\sum_{\alpha\beta}\phi_{a}(x_{i},u_{\alpha})\phi_{a}(x_{j},u_{\alpha})\phi_{a}(x_{i},u_{\beta})\phi_{a}(x_{j},u_{\beta}),
\]
\[
\widehat{\mathcal{H}_{0}}(\phi_{b},\mathcal{S}_{\mathcal{X}},\mathcal{F}_{\mathcal{V}})=\frac{1}{P^{2}Q_{b}^{2}}\sum_{ij}\sum_{\alpha\beta}\phi_{b}(x_{i},v_{\alpha})\phi_{b}(x_{j},v_{\alpha})\phi_{b}(x_{i},v_{\beta})\phi_{b}(x_{j},v_{\beta})
\]

Here, we assume the naive CKA is computed after $N$ trials, across
which the empirical means of the HSIC are computed::

\[
\widehat{\text{CKA}}_{0}(\phi_{a},\phi_{b},\rho_{\mathcal{X}},\rho_{\mathcal{U}},\rho_{\mathcal{V}})=\frac{\sum_{l=1}^{N}\widehat{\mathcal{H}_{0}}(\phi_{a},\phi_{b},\mathcal{S}_{\mathcal{X}}^{(l)},\mathcal{F}_{\mathcal{U}}^{(l)},\mathcal{F}_{\mathcal{V}}^{(l)})}{\sqrt{\sum_{l=1}^{N}\widehat{\mathcal{H}_{0}}(\phi_{a},\mathcal{S}_{\mathcal{X}}^{(l)},\mathcal{F}_{\mathcal{U}}^{(l)})\sum_{l=1}^{N}\widehat{\mathcal{H}_{0}}(\phi_{b},\mathcal{S}_{\mathcal{X}}^{(l)},\mathcal{F}_{\mathcal{V}}^{(l)})}}
\]
We assume $N$ is large such that all three $\widehat{\mathcal{H}}$ empirical averages
have small variance of order $\mathcal{O}\left(\frac{1}{N}\right)$.
In this limit, the following approximation is valid:

\[
\left\langle \widehat{\text{CKA}}_{0}(\phi_{a},\phi_{b},\mathcal{S}_{\mathcal{X}},\mathcal{F}_{\mathcal{U}},\mathcal{F}_{\mathcal{V}})\right\rangle \approx\frac{\left\langle \widehat{\mathcal{H}_{0}}(\phi_{a},\phi_{b},\mathcal{S}_{\mathcal{X}}^{(l)},\mathcal{F}_{\mathcal{U}}^{(l)},\mathcal{F}_{\mathcal{V}}^{(l)})\right\rangle }{\sqrt{\left\langle \widehat{\mathcal{H}_{0}}(\phi_{a},\mathcal{S}_{\mathcal{X}}^{(l)},\mathcal{F}_{\mathcal{U}}^{(l)})\right\rangle \left\langle \widehat{\mathcal{H}_{0}}(\phi_{b},\mathcal{S}_{\mathcal{X}}^{(l)},\mathcal{F}_{\mathcal{V}}^{(l)})\right\rangle }}.
\]
This approximation is also valid when $P$, $Q_{a}$, and $Q_{b}$
are all large. We empirically observe that the expected value of $\widehat{\text{CKA}}_{0}$
obtained via this approximation still accurately predict the expected
values of $\widehat{\text{CKA}}_{0}$ empirically computed even with
$N=1$ and often even in addition to small $P$, $Q_{a}$, and $Q_{b}$.
This allows us to understand the bias of $\widehat{\text{CKA}}_{0}$
contributed from the biases of $\widehat{\mathcal{H}}_{0}$, isolated
from the bias contributed by taking products, inverse, and square
root of the $\widehat{\mathcal{H}}_{0}$ estimates, which unnecessarily
complicates the analysis.

First, let us take the expected value of the $\widehat{\mathcal{H}}_{0}$
in the numerator:
\begin{align}
\left\langle \widehat{\mathcal{H}_{0}}(\phi_{a},\phi_{b},\mathcal{S}_{\mathcal{X}},\mathcal{F}_{\mathcal{U}},\mathcal{F}_{\mathcal{V}})\right\rangle =& \frac{1}{P^{2}Q_{a}Q_{b}}\sum_{ij}\sum_{\alpha}\sum_{\beta}\left\langle \phi_{a}(x_{i},u_{\alpha})\phi_{a}(x_{j},u_{\alpha})\phi_{b}(x_{i},v_{\beta})\phi_{b}(x_{j},v_{\beta})\right\rangle\\     
=&\left\langle k_{a}(x,y)k_{b}(x,y)\right\rangle _{x,y}+\frac{1}{P}\left(\left\langle k_{a}(x,x)k_{b}(x,x)\right\rangle _{x}-\left\langle k_{a}(x,y)k_{b}(x,y)\right\rangle _{x,y}\right)\\
\label{eqn:numer_bias}
=&\left\langle k_{a}(x,y)k_{b}(x,y)\right\rangle _{x,y}\left(1+\frac{1}{P}\left(\zeta_{ab}-1\right)\right)
\end{align}

where we have introduced a new variable $\zeta_{ab}\coloneqq\frac{\left\langle k_{a}(x,x)k_{b}(x,x)\right\rangle _{x}}{\left\langle k_{a}(x,y)k_{b}(x,y)\right\rangle _{x,y}}$
that is sensitive to the alignment of the representation. Note that the expected value of the stimulus-corrected estimator $\widehat{\mathcal{H}_{S}}(\phi_{a},\phi_{b},\mathcal{S}_{\mathcal{X}},\mathcal{F}_{\mathcal{U}},\mathcal{F}_{\mathcal{V}})$ can be obtained by taking $P\to\infty$ limit in \cref{eqn:numer_bias}, which simply $\left\langle k_{a}(x,y)k_{b}(x,y)\right\rangle$, which means that the numerator $\widehat{\mathcal{H}_{S}}$ is unbiased in this problem setup. However, in an alternative problem setup, such as the trial-to-trial CKA, the numerator $\widehat{\mathcal{H}_{S}}$ is still biased, since the neurons are identical across (a) and (b), i.e. $u_\alpha=v_\alpha$. The bias of the numerator $\widehat{\mathcal{H}_{S}}$ in this alternative problem setup is similar to the biases in the denominator $\widehat{\mathcal{H}_{S}}$'s of both problem setup.

Next, let us take the expected value of one of the $\widehat{\mathcal{H}}_{0}$'s
in the denominator:

\[
\left\langle \widehat{\mathcal{H}_{0}}(\phi_{a},\mathcal{S}_{\mathcal{X}},\mathcal{F}_{\mathcal{U}})\right\rangle =\frac{1}{P^{2}Q_{a}^{2}}\sum_{ij}\sum_{\alpha\beta}\left\langle \phi_{a}(x_{i},u_{\alpha})\phi_{a}(x_{j},u_{\alpha})\phi_{a}(x_{i},u_{\beta})\phi_{a}(x_{j},u_{\beta})\right\rangle 
\]

\begin{multline}
=\left\langle k_{a}(x,y)^{2}\right\rangle -\frac{1}{P}\left(\left\langle k_{a}(x,y)^{2}\right\rangle -\left\langle k_{a}(x,x)^{2}\right\rangle \right)-\frac{1}{Q_{a}}\left(\left\langle k_{a}(x,y)^{2}\right\rangle -\left\langle \tilde{k}_{a}(w,w)^{2}\right\rangle \right)+\\
\frac{1}{PQ_{a}}\left(\left\langle k_{a}(x,y)^{2}\right\rangle -\left\langle k_{a}(x,x)^{2}\right\rangle -\left\langle \tilde{k}_{a}(w,w)^{2}\right\rangle +\left\langle \phi_{a}(x,w)^{4}\right\rangle \right)    
\end{multline}

\[
=\left\langle k_{a}(x,y)^{2}\right\rangle \left[1+\frac{1}{P}\left(\frac{\gamma_{a}}{\psi_{a}}-1\right)+\frac{1}{Q_{a}}\left(\frac{\gamma_{a}}{\tilde{\psi}_{a}}-1\right)-\frac{1}{PQ_{a}}\left(\frac{\gamma_{a}}{\psi_{a}}+\frac{\gamma_{a}}{\tilde{\psi}_{a}}-\frac{\gamma_{a}}{\rho_{a}}-1\right)\right]
\]
where we have introduced new variables $\gamma_{a}=\frac{\left\langle k_{a}(x,x)\right\rangle ^{2}}{\left\langle k_{a}(x,y)^{2}\right\rangle }$,
$\psi_{a}\coloneqq\frac{\left\langle k_{a}(x,x)\right\rangle ^{2}}{\left\langle k_{a}(x,x)^{2}\right\rangle }$,
$\tilde{\psi}_{a}\coloneqq\frac{\left\langle \tilde{k}_{a}(w,w)\right\rangle ^{2}}{\left\langle \tilde{k}_{a}(w,w)^{2}\right\rangle }$,
and $\rho_{a}=\frac{\left\langle \phi_{a}(x,w)^{2}\right\rangle ^{2}}{\left\langle \phi_{a}(x,w)^{4}\right\rangle }$.
Each of them is participation ratio (PR), i.e. effective/soft count,
of some quantities. For discrete quantities, let us define PR as
\[
\frac{\left(\sum_{i}a_{i}\right)^{2}}{\sum_{i}a_{i}^{2}}.
\]
It is easy to see that, if $N$ number of $a_{i}$'s take value $1$
and the rest $0$, then PR is $N$, indicating it is a soft count
of non-zero $a_{i}$. The continuous version is

\[
\frac{\left(\int d\mu(t)\,f(t)\right)^{2}}{\int d\mu(t)\,f(t)^{2}},
\]
which we call PR of $f$ w.r.t. to $\mu$. With these definitions,
we can interpret$\gamma_{a}$ as the PR of the eigenvalues of $T_{k_{a}}$,
i.e. intrinsic dimensionality of $T_{k}$, $\psi_{a}$ as the PR of
$\int d\rho_{\mathcal{U}}(w)\:\phi_{a}(\cdot,w)^{2}$ w.r.t. $\rho_{\mathcal{X}}$,
$\tilde{\psi}_{a}$ as the PR of $\int d\rho_{\mathcal{X}}(x)\:\phi_{a}(x,\cdot)^{2}$
w.r.t. $\rho_{\mathcal{U}}$, and $\rho_{a}$ as the PR of $\phi_{a}^{2}$
w.r.t. $\rho_{\mathcal{X}}\otimes\rho_{\mathcal{U}}$. Suppose $\Phi_{\infty}^{(a)}$
is the measurement matrix in the limit of infinite stimulus and neuron
samples, and $\mathbf{K}_{\infty}^{(a)}$ is the corresponding Gram
matrix, equivalent to $T_{k_{a}}$. Then we can loosely say $\gamma_{a}$
is the intrinsic dimensionality of $\mathbf{K}_{\infty}$, $\psi_{a}$
is the effective number of rows of $\Phi_{\infty}^{(a)}$ with non-zero
lengths, $\tilde{\psi_{a}}$ is the effective number of columns of
$\Phi_{\infty}^{(a)}$ with non-zero lengths, and $\rho_{a}$ is the
effective number of non-zero entries in $\Phi_{\infty}^{(a)}$.

Then, with the expected values of $\widehat{\mathcal{H}}_{0}$ derived
above the expected value of $\widehat{\text{CKA}}_{0}$ can be approximated
as

\begin{equation}\label{Seq:o_bias}
\left\langle \widehat{\text{CKA}}_{0}(\phi_{a},\phi_{b},\mathcal{S}_{\mathcal{X}},\mathcal{F}_{\mathcal{U}},\mathcal{F}_{\mathcal{V}})\right\rangle 
\approx
\frac{\left(1+\frac{1}{P}\left(\zeta_{ab}-1\right)\right)\text{CKA}}{\sqrt{\left(1+\frac{\frac{\gamma_{a}}{\psi_{a}}-1}{P}+\frac{\frac{\gamma_{a}}{\tilde{\psi}_{a}}-1}{Q_{a}}-\frac{\frac{\gamma_{a}}{\psi_{a}}+\frac{\gamma_{a}}{\tilde{\psi}_{a}}-\frac{\gamma_{a}}{\rho_{a}}-1}{PQ_{a}}\right)\left(1+\frac{\frac{\gamma_{b}}{\psi_{b}}-1}{P}+\frac{\frac{\gamma_{b}}{\tilde{\psi}_{b}}-1}{Q_{b}}-\frac{\frac{\gamma_{b}}{\psi_{b}}+\frac{\gamma_{b}}{\tilde{\psi}_{b}}-\frac{\gamma_{b}}{\rho_{b}}-1}{PQ_{b}}\right)}}
\end{equation}

where $\text{CKA}\coloneqq\frac{\left\langle k_{a}(x,y)k_{b}(x,y)\right\rangle }{\sqrt{\left\langle k_{a}(x,y)^{2}\right\rangle \left\langle k_{b}(x,y)^{2}\right\rangle }}$
is the true CKA. Taking $P\to\infty$, we obtain the expected value
of $\widehat{\text{CKA}}_{S}(\phi_{a},\phi_{b},\mathcal{S}_{\mathcal{X}},\mathcal{F}_{\mathcal{U}},\mathcal{F}_{\mathcal{V}})$
estimator:

\[
\left\langle \widehat{\text{CKA}}_{S}(\phi_{a},\phi_{b},\mathcal{S}_{\mathcal{X}},\mathcal{F}_{\mathcal{U}},\mathcal{F}_{\mathcal{V}})\right\rangle 
\approx
\frac{\text{CKA}}{\sqrt{\left(1+\frac{\frac{\gamma_{a}}{\tilde{\psi}_{a}}-1}{Q_{a}}\right)\left(1+\frac{\frac{\gamma_{b}}{\tilde{\psi}_{b}}-1}{Q_{b}}\right)}}
\]
If $\int d\rho_{\mathcal{X}}(x)\:\phi_{a}(x,\cdot)^{2}$ and $\int d\rho_{\mathcal{X}}(x)\:\phi_{b}(x,\cdot)^{2}$
are constant, i.e. the norm of the activation is normalized for each
neuron, then $\tilde{\psi}_{a}=\tilde{\psi}_{b}=1$, which gives

\[
\left\langle \widehat{\text{CKA}}_{S}(\phi_{a},\phi_{b},\mathcal{S}_{\mathcal{X}},\mathcal{F}_{\mathcal{U}},\mathcal{F}_{\mathcal{V}})\right\rangle 
\approx
\frac{\text{CKA}}{\sqrt{\left(1+\frac{\gamma_{a}-1}{Q_{a}}\right)\left(1+\frac{\gamma_{b}-1}{Q_{b}}\right)}}
\]

\subsection{Analyzing the overall \texorpdfstring{$\widehat{\cH}$}{H}-based CKA estimators}\label{S:hardbias}

All CKA estimators based on $\widehat{\mathcal{H}}$, including ours, have
bias coming from taking the product of $\widehat{\mathcal{H}}$'s that are
correlated to each other, taking inverse, and taking the square root.
As a simple analog of CKA estimator, consider the following problem.
We have fixed quantities $X$, $A$, and $B$, and corresponding potentially
biased estimators $x$, $a$, and $b$, respectively. Here, $x$,
$a$, and $b$ are correlated. We wish to
study the properties of the estimator 
\[
T=\frac{x}{\sqrt{ab}},
\]
as an estimator for 
\[
\theta=\frac{X}{\sqrt{AB}}.
\]
Here, the fixed quantities $X$, $A$, and $B$ correspond to the
ground truth $\mathcal{H}$ values, and $x$, $a$, and $b$ correspond
to the $\widehat{\mathcal{H}}$ estimators.

We denote 
\[
\begin{aligned}\delta_{x} & =E[x]-X,\quad\delta_{a}=E[a]-A,\quad\delta_{b}=E[b]-B,\\[1mm]
\sigma_{x}^{2} & =\text{Var}(x),\quad\sigma_{a}^{2}=\text{Var}(a),\quad\sigma_{b}^{2}=\text{Var}(b),\\[1mm]
\sigma_{xa} & =\text{Cov}(x,a),\quad\sigma_{xb}=\text{Cov}(x,b),\quad\sigma_{ab}=\text{Cov}(a,b).
\end{aligned}
\]
We also define the deviations: 
\[
\Delta x=x-X,\quad\Delta a=a-A,\quad\Delta b=b-B,
\]
with 
\[
E[\Delta x]=\delta_{x},\quad E[\Delta a]=\delta_{a},\quad E[\Delta b]=\delta_{b}.
\]

We consider the function 
\[
f(x,a,b)=\frac{x}{\sqrt{ab}},
\]
and expand it about the point $(X,A,B)$ to second order: 

\[
f(x,a,b)\approx f(X,A,B)+f_{x}\Delta x+f_{a}\Delta a+f_{b}\Delta b+\frac{1}{2}\left[f_{xx}(\Delta x)^{2}+f_{aa}(\Delta a)^{2}+f_{bb}(\Delta b)^{2}+2f_{xa}\Delta x\Delta a+2f_{xb}\Delta x\Delta b+2f_{ab}\Delta a\Delta b\right]
\]
with all derivatives evaluated at $(X,A,B)$.

The zeroth-order term:

\[
f(X,A,B)=\frac{X}{\sqrt{AB}}.
\]

The first order derivatives:
\[
f_{x}(X,A,B)=\frac{1}{\sqrt{AB}},\quad f_{a}(X,A,B)=-\frac{X}{2A\sqrt{AB}},\quad f_{b}(X,A,B)=-\frac{X}{2B\sqrt{AB}}.
\]

Next, the second-order derivatives:

\[
f_{xx}(X,A,B)=0,\quad f_{xa}(X,A,B)=-\frac{1}{2A\sqrt{AB}},\quad f_{xb}(X,A,B)=-\frac{1}{2B\sqrt{AB}},\quad
\]

\[
f_{aa}(X,A,B)=\frac{3X}{4A^{2}\sqrt{AB}},\quad f_{bb}(X,A,B)=\frac{3X}{4B^{2}\sqrt{AB}},\quad f_{ab}(X,A,B)=\frac{X}{4AB\sqrt{AB}}.
\]
Plugging them into our expansion, we get

\[
\frac{x}{\sqrt{ab}}\approx\frac{X}{\sqrt{AB}}\left(1+\frac{\Delta x}{X}-\frac{1}{2}\frac{\Delta a}{A}-\frac{1}{2}\frac{\Delta b}{B}+\frac{3}{8}\frac{(\Delta a)^{2}}{A^{2}}+\frac{3}{8}\frac{(\Delta b)^{2}}{B^{2}}-\frac{1}{2}\frac{\Delta x\Delta a}{XA}-\frac{1}{2}\frac{\Delta x\Delta b}{XB}+\frac{1}{4}\frac{\Delta a\Delta b}{AB}\right)
\]

Now let us take the expected value of the above. We use 
\[
\begin{aligned}E[\Delta x] & =\delta_{x},\quad E[\Delta a]=\delta_{a},\quad E[\Delta b]=\delta_{b},\\[1mm]
E[(\Delta x)^{2}] & =\sigma_{x}^{2}+\delta_{x}^{2},\quad E[(\Delta a)^{2}]=\sigma_{a}^{2}+\delta_{a}^{2},\quad E[(\Delta b)^{2}]=\sigma_{b}^{2}+\delta_{b}^{2},\\[1mm]
E[\Delta x\Delta a] & =\sigma_{xa}+\delta_{x}\delta_{a},\quad E[\Delta x\Delta b]=\sigma_{xb}+\delta_{x}\delta_{b},\quad E[\Delta a\,\Delta b]=\sigma_{ab}+\delta_{a}\delta_{b}.
\end{aligned}
\]
With these definitions, we get the following as the expected value
up to the second order approximation:
\[
E\left[\frac{x}{\sqrt{ab}}\right]\approx\frac{X}{\sqrt{AB}}\left(1+\frac{\delta_{x}}{X}-\frac{1}{2}\frac{\delta_{a}}{A}-\frac{1}{2}\frac{\delta_{b}}{B}+\frac{3}{8}\frac{\sigma_{a}^{2}+\delta_{a}^{2}}{A^{2}}+\frac{3}{8}\frac{\sigma_{b}^{2}+\delta_{b}^{2}}{B^{2}}-\frac{1}{2}\frac{\sigma_{xa}+\delta_{x}\delta_{a}}{XA}-\frac{1}{2}\frac{\sigma_{xb}+\delta_{x}\delta_{b}}{XB}+\frac{1}{4}\frac{\sigma_{ab}+\delta_{a}\delta_{b}}{AB}\right).
\]
Therefore, the bias is given by

\[
E\left[\frac{x}{\sqrt{ab}}\right]-\frac{X}{\sqrt{AB}}\approx\frac{X}{\sqrt{AB}}\left(\frac{\delta_{x}}{X}-\frac{1}{2}\frac{\delta_{a}}{A}-\frac{1}{2}\frac{\delta_{b}}{B}+\frac{3}{8}\frac{\sigma_{a}^{2}+\delta_{a}^{2}}{A^{2}}+\frac{3}{8}\frac{\sigma_{b}^{2}+\delta_{b}^{2}}{B^{2}}-\frac{1}{2}\frac{\sigma_{xa}+\delta_{x}\delta_{a}}{XA}-\frac{1}{2}\frac{\sigma_{xb}+\delta_{x}\delta_{b}}{XB}+\frac{1}{4}\frac{\sigma_{ab}+\delta_{a}\delta_{b}}{AB}\right).
\]

Now let us put this back into the perspective of the CKA estimation.
In that context, $\delta_{x}$ is the bias of an $\mathcal{H}$-estimator
for the numerator, whereas $\delta_{a}$ and $\delta_{b}$ are the
bias of the $\mathcal{H}$-estimators in the numerator of CKA estimator.
The variances $\sigma_{x}^{2}$, $\sigma_{a}^{2}$, and $\sigma_{b}^{2}$
are the variances of the corresponding $\mathcal{H}$-estimators.
Again, $X$, $A$, $B$ corresponds to the true $\mathcal{H}$-values
in the true CKA. In our estimator $\widehat{\text{CKA}}_{C}$, the
$\widehat{\mathcal{H}}_{C}$ biases $\delta_{x}$, $\delta_{a}$, and
$\delta_{b}$ are zero, substantially reducing the bias in $\widehat{\text{CKA}}_{C}$.
However, in $\widehat{\text{CKA}}_{0}$ and $\widehat{\text{CKA}}_{S}$,
their $\mathcal{H}$ estimators have non-zero biases $\delta_{x},\delta_{a},\delta_{b}>0$
of order $\mathcal{O}\left(\frac{1}{P}+\frac{1}{Q}\right)$ (except,
for some setups, $\delta_{x}$ is zero in $\widehat{\text{CKA}}_{S}$),
adding even more bias to the CKA estimations. Note that all $\mathcal{H}$
estimators have variance ($\sigma_{x}^{2}$, $\sigma_{a}^{2}$, $\sigma_{b}^{2}$,
$\sigma_{xa}$, $\sigma_{xb}$, and $\sigma_{ab}$) of order $\mathcal{O}\left(\frac{1}{P}+\frac{1}{Q}\right)$.
Therefore, as long as CKA estimation is built based on the $\mathcal{H}$
estimators, the overall bias of CKA is of order $\mathcal{O}\left(\frac{1}{P}+\frac{1}{Q}\right)$.

In a scenario where we have two systems (a) and (b) and we aim to
compare the representations in (a) and (b) for a single input distribution.
However, for each trial, one uses a unique set of inputs and observes
a unique set of neurons independently drawn from distributions. Then,
instead of taking the empirical average of $\widehat{\text{CKA}}$,
one can take the empirical averages of $\widehat{\mathcal{H}}$ and
use that average to compute CKA. If there are $N$ trials, these empirical
averages of $\widehat{\mathcal{H}}$ have the variance of order $\mathcal{O}\left(\frac{1}{N}\left(\frac{1}{P}+\frac{1}{Q}\right)\right)$.
This situation corresponds to when $N$ number of labs use distinct
individual animals to study V1 and they use distinct sets of natural
images. Alternatively, if one aims to compute trial-to-trial CKA and
there are $M$ trials, then by performing all pairwise comparisons,
one can reduce the bias of CKA estimators significantly since $N=\binom{M}{2}$.
In summary, having multiple trials allows our estimator $\widehat{\text{CKA}}_{C}$
to have bias of $\mathcal{O}\left(\frac{1}{N}\left(\frac{1}{P}+\frac{1}{Q}\right)\right)$,
while the other estimators $\widehat{\text{CKA}}_{0}$ and $\widehat{\text{CKA}}_{S}$
still has bias of $\mathcal{O}\left(\frac{1}{P}+\frac{1}{Q}\right)$.

\section{Additional results on neural data}
In this section, we present the results on the second monkey subject. Overall, our observations are consistent across the two monkeys.

\subsection{Brain-to-model alignment}\label{S:B2M}
\begin{figure}[h]
    \centering
    \includegraphics[width=\textwidth]{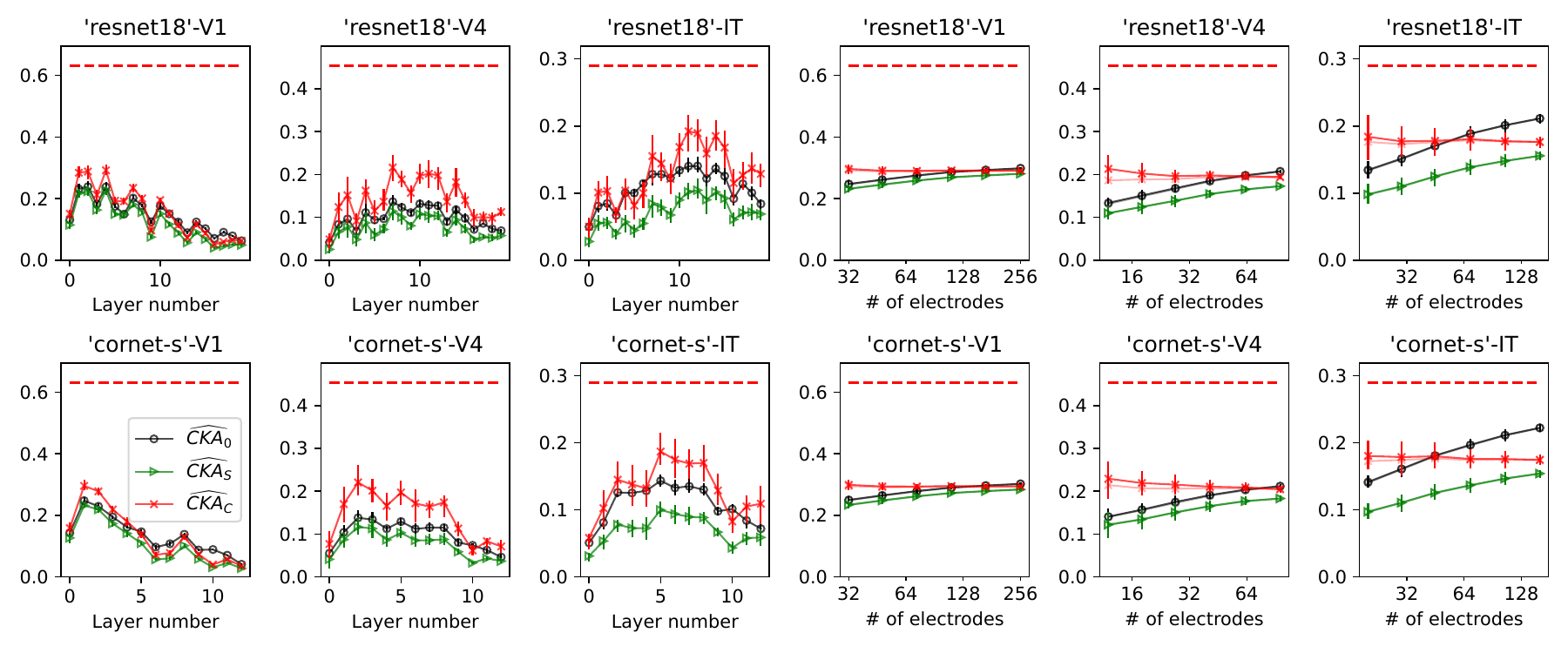}
    \caption{Comparison of the brain regions (V1,V4, and IT) and models (ResNet18 and CorNet-s) using CKA estimators. The left three columns show CKAs between each brain region and all model layers. The number of electrodes ($Q$) is downsampled to $1/16$ of all available electrodes in the dataset for each brain region.
    The right three columns show CKA between a brain region and the layer that is the most similar to the region. The number of electrodes ($Q$) varies from the factor of $1/16$ to $1$ (of all available electrodes) along the x-axis for each region. $P=1,000$ stimuli are used in all plots. The horizontal dotted line is the brain-to-brain $\widehat{\text{CKA}}_{C}$ across animals for each brain region from \cref{fig:b2b}b. The darker lines are $\widehat{\text{CKA}}$'s and the lighter lines are $\widehat{\text{CKA}^N}$'s.}
    \label{Sfig:monkey1_model}
\end{figure}
The CKA estimates between CNNs and the second monkey is shown in \cref{Sfig:monkey1_model}.

\subsection{Object disentanglement}\label{S:OD}

\begin{figure}[!t]
    \centering
    \includegraphics[width=\textwidth]{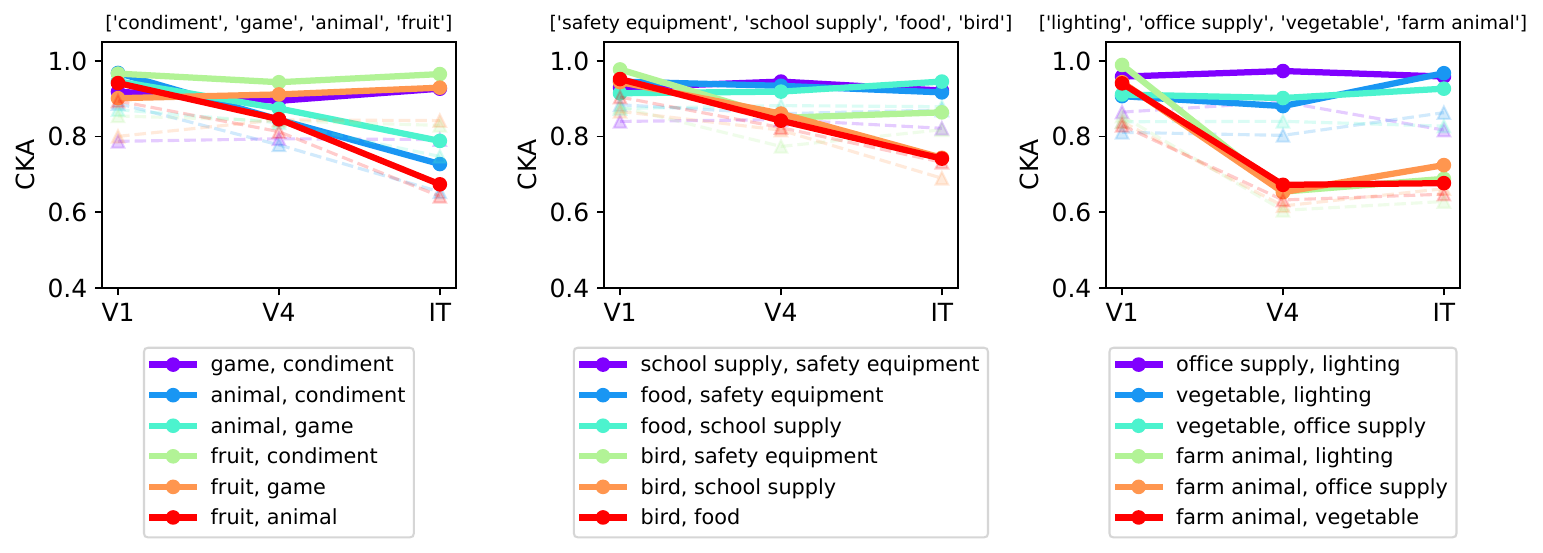}
    \caption{Quantifying semantic disentanglement in the brain with CKA. Solid line represents $\widehat{\text{CKA}}_{C}$, whereas dotted line represents $\widehat{\text{CKA}}_{S}$. Three separate groups of object images were prepared, and all pairwise comparisons were performed in each group.}
    \label{Sfig:obj1}
\end{figure}
The CKA estimates object disentanglement in the second monkey is shown in \cref{Sfig:obj1}.

\section{Code availability}\label{S:code}
The code for the estimators and generating all figures is available in \url{https://github.com/badooki/CKA/}.

\end{document}